# The Small World of Osteocytes:
# Connectomics of the Lacuno-Canalicular Network in Bone

Philip Kollmannsberger[1,2,*], Michael Kerschnitzki[1,3], Felix Repp[1], Wolfgang Wagermaier[1], Richard Weinkamer[1], Peter Fratzl[1]

[1]Max Planck Institute of Colloids and Interfaces, Department of Biomaterials, Potsdam, Germany

[2]ETH Zurich, Laboratory of Applied Mechanobiology, Department of Health Sciences and Technology, Zurich, Switzerland

[3]Weizmann Institute of Science, Dept. of Structural Biology, Rehovot, Israel

[*] current address: Center for Computational and Theoretical Biology, University of Würzburg, Würzburg, Germany

**Abstract**

Osteocytes and their cell processes reside in a large, interconnected network of voids pervading the mineralized bone matrix of most vertebrates. This osteocyte lacuno-canalicular network (OLCN) is believed to play important roles in mechanosensing, mineral homeostasis, and for the mechanical properties of bone. While the extracellular matrix structure of bone is extensively studied on ultrastructural and macroscopic scales, there is a lack of quantitative knowledge on how the cellular network is organized. Using a recently introduced imaging and quantification approach, we analyze the OLCN in different bone types from mouse and sheep that exhibit different degrees of structural organization not only of the cell network but also of the fibrous matrix deposited by the cells. We define a number of robust, quantitative measures that are derived from the theory of complex networks. These measures enable us to gain insights into how efficient the network is organized with regard to intercellular transport and communication. Our analysis shows that the cell network in regularly organized, slow-growing bone tissue from sheep is less connected, but more efficiently organized compared to irregular and fast-growing bone tissue from mice. On the level of statistical topological properties (edges per node, edge length and degree distribution), both network types are indistinguishable, highlighting that despite pronounced differences at the tissue level, the topological architecture of the osteocyte canalicular network at the subcellular level may be independent of species and bone type. Our results suggest a universal mechanism underlying the self-organization of individual cells into a large, interconnected network during bone formation and mineralization.

# 1 Introduction

Network structures are ubiquitous in Nature and often fulfil important functions in transport and signal processing. In humans, examples include the airways (Metzger, Klein et al. 2008), the vasculature (Blinder, Tsai et al. 2013, Pries and Secomb 2014) and the nervous system (Eguiluz, Chialvo et al. 2005, Bullmore and Sporns 2009). The organization into networks can already occur on the level of the individual cells, with neurons using their extended dendrites and synapses to connect with other cells as the prototypical example (Cajal and May 1928, Helmstaedter, Briggman et al. 2013, Takemura, Bharioke et al. 2013). It has also been known for a long time that the mineralized bone tissue of most vertebrates is densely populated by cells called osteocytes. These cells are embedded in the extracellular bone matrix during bone deposition, and are linked with each other and with blood vessels through a highly interconnected network of cell processes (Bonewald 2011) in appearance and size similar to the neuronal system (Buenzli and Sims 2015). Osteocyte bodies and their processes reside in hollow lacunae and narrow canals termed canaliculi, respectively, that together comprise the osteocyte lacuno-canalicular network (OLCN) (Franz-Odendaal, Hall et al. 2006). The functional relevance of this network and the details of its architecture are still under debate, despite a large number of recent studies in this field (Asada, Katayama et al. 2013, Thi, Suadicani et al. 2013, Hesse, Varga et al. 2015, Milovanovic, Zimmermann et al. 2015, Sano, Kikuta et al. 2015, Nango, Kubota et al. 2016).

One assumed function of osteocytes is to orchestrate the process of structural adaptation and material renewal in bone. These processes are thought to be mechanically controlled, with the result that new bone is added where mechanically needed and resorbed where local loading is low. The role of the osteocytes is to locally initiate bone (re)modeling by generating biochemical signals that activate precursor cells in response to mechanical stimuli (Nakashima, Hayashi et al. 2011). In this case, the role of the OLCN would be that of a signal amplifier: small deformations or cracks within the bone tissue are sensed by osteocyte processes and transduced into a biochemical response that is then spatially integrated and triggers the recruitment of osteoblast and osteoclast precursor cells (Wang, McNamara et al. 2007). Besides this putative mechano-sensory role, osteocytes are suspected to be able to locally remodel the bone matrix that surrounds them (osteocytic osteolysis) (Belanger 1969,

Teti and Zallone 2009). This would enable them to participate in the calcium and phosphate metabolism of the body by making the mineral reservoir deep in the bone tissue accessible through the network of canaliculi. The ability of osteocytes to secrete endocrine factors involved in mineral homeostasis strongly supports this possibility (Dallas, Prideaux et al. 2013). For both, mechanosensing and mineral exchange, the efficiency of the network in distributing molecules and signals between cells and throughout the matrix plays a central role. Finally, the function of the canalicular network for the osteocytes is to supply nutrients and remove waste by maintaining contact to blood vessels (Bonewald 2011).

The efficiency of real-world networks such as roadmaps, metabolic networks, or the internet can be assessed using methods and measures derived from graph theory (Boccaletti, Latora et al. 2006). By representing these diverse systems as a collection of edges connected by nodes, their structural and topological properties can be analyzed and compared. Common features of real-world networks include exponential or scale-free degree distributions, high clustering coefficients, and hierarchical organization. In contrast to random graphs, empirical networks often exhibit small-world properties such that most points in the network can be reached from anywhere by a small number of steps, which is very efficient for transport and communication. The application of complex network theory to biological networks resulted in many important insights into the structure and function of these networks (Bullmore and Sporns 2009, Pries and Secomb 2014).

To date, such a characterization could not be carried out for the osteocyte network in bone, in part due to a lack of structural data that provide sufficient resolution and at the same time span large enough volumes. Direct imaging of osteocytes and their processes in living bone is difficult due to the limited accessibility and transparency of the bone matrix; therefore, most techniques for imaging the osteocyte network rely on imaging the OLCN in bone samples after cells and soft tissue have been removed. Mainly, three classes of imaging modalities have been reported, based on X-ray micro computed tomography (μCT), scanning electron microscopy (SEM), or confocal laser scanning microscopy (CLSM). While conventional μCT scanners can image individual lacunae including their shape and orientation, phase-resolved synchrotron X-ray tomography has recently been applied to derive 3D images of the network of canaliculi around a single osteocyte at a few nm resolution ((Dierolf, Menzel et al. 2010, Hesse, Varga et al. 2015). SEM can only image surfaces due to the low penetration depth of electrons in dense tissue, but two approaches have been reported to derive 3D data of the OLCN using SEM. In one approach, the hollow spaces in bone are filled with a resin, and the surrounding bone tissue is then etched away, exposing the resin cast and hence the structure of the OLCN as a projection (Pazzaglia and Congiu 2013). The other approach is FIB-SEM slice-and-view, where the surface is repeatedly imaged with SEM and milled using a focused ion beam. This variant produced the most detailed and accurate 3D images of the OLCN (Schneider, Meier et al. 2011) and its embedment into the collagen matrix (Reznikov, Shahar et al. 2014) to date, yet both SEM-based techniques inevitably result in the destruction of the sample. The third method, CLSM, requires fluorescent staining of the network while the sample itself is preserved (Kerschnitzki, Wagermaier et al. 2011). The resolution of CLSM is limited by the optical diffraction limit and is intrinsically anisotropic and depth-dependent, and therefore, in contrast to XRPT and FIB-SEM, cannot accurately resolve the diameter of individual canaliculi.

We previously showed that by image processing, the network can be reconstructed from CLSM image stacks as long as the resolution is sufficient to separate neighboring canaliculi (Kerschnitzki, Kollmannsberger et al. 2013). This study revealed a relationship between the proximity of the bone matrix to canaliculi and the nanoscale properties of the mineral crystals in the matrix, suggesting active participation of osteocytes in bone remodeling and mineral homeostasis. Here, we present an improvement of this imaging and analysis technique and use it to quantify the architecture of the osteocyte network in different bone types from mouse and sheep. We use fibrolamellar bone from sheep, which grows slower but in a more controlled way on top of a template surface (Ferretti, Palumbo et al. 2002, Liu, Manjubala et al. 2010, Kerschnitzki, Wagermaier et al. 2011), and woven bone from mice, which grows faster but in the absence of an organizing surface. Based on these data, we report the first detailed characterization of its topological and statistical properties on the subcellular level using measures from complex network theory. The OLCN reconstructed from CLSM image stacks is described as a network graph, i.e. by the locations of nodes and edges, and the adjacency matrix, which contains information on how nodes are connected through edges. This makes it possible to calculate generic statistical and topological properties such as edge length and degree distributions, and relationships between these properties. With this proof-of-principle approach, we aim to address the question whether the known differences in tissue organization and appearance of the OLCN in different types of bone from different species are reflected in the organization of the network of cell processes, or if the network at the cellular and subcellular level is independent of bone type and level of organization. By developing a quantitative characterization of the topological and statistical properties of the OLCN in different bone types, we hope to learn more about what purpose this network may have evolved for, how its formation is controlled, and how these network measures relate to bone material quality during physiological development or pathological conditions of age, disease and pharmaceutical intervention.

## 2 Materials and Methods

### 2.1 *Sample preparation and imaging*

The workflow from sample preparation to image analysis and quantification is summarized in Fig. 1a. Bone samples, sample preparation, and staining with Rhodamine-6G were the same as previously described (Kerschnitzki, Wagermaier et al. 2011, Kerschnitzki, Kollmannsberger et al. 2013). We used fibrolamellar ovine bone from the mid-diaphysis of the femur of a 5 year old sheep as well as murine bone from the mid-diaphysis of the femur from a 12 month-old mouse, without initial aldehyde fixation. Bone samples were cut either in cross sections or in longitudinal sections with an initial thickness of 200 microns and polished from both sides in an automatic polisher (Logitech PM5, Logitech Ltd., Glasglow, UK) until a final thickness of 80 microns. Samples were kept wet throughout the whole preparation process. Until further usage, samples were wrapped in cotton gauze soaked in phosphate buffered saline (PBS) solution (Sigma Aldrich GmbH, Germany) and stored at 4 °C. For each bone type and orientation, five different fields of view were imaged, amounting to 20 image stacks with 51 images per stack in total. Image acquisition was done on a Leica DM IRBE confocal microscope (Leica Microsystems, Wetzlar) using a 100x oil immersion objective (NA = 1.4) using a voxel size of ~ 0.2 µm x 0.2 µm x 0.2 µm. Total volume size was 512 x 512 x 51 voxels or 100 µm x 100 µm x 10 µm.

### 2.2 *Image processing*

Raw image stacks were Gaussian filtered in 3D ($\sigma = 0.65$ voxels) to remove high-frequency noise, and a *top hat* filter with a disk-shaped kernel of radius 25 pixels was applied to each image slice separately. Top-hat filtering subtracts the eroded and dilated image from the original image, thereby removing intensity gradients on length scales larger than the kernel in the image plane before applying a global intensity threshold. The threshold for separating foreground and background voxels was defined as the intensity where the number of non-connected objects in the binarized image was minimal (Kerschnitzki, Kollmannsberger et al. 2013). Smaller values would result in an increase of the number of objects due to noise, whereas higher values would lead to fragmentation of the network, again increasing the number of objects. This method was applied to a number of representative stacks to determine the value for the threshold, which was then fixed and applied globally to all pre-processed stacks. All objects in the resulting binary volumes smaller than 10 voxels were considered as noise and removed. Finally, the image was morphologically closed to fill gaps, and all isolated background components (e.g. voids inside lacunae resulting from incomplete staining) were filled.

## *2.3 Cell lacunae detection and skeletonization*

Cell lacunae and other large structures (e.g. blood vessels) were detected using a modified 3D watershed algorithm. All objects more than 7 voxels away from the surface and more than 100 voxels in size were defined as seed points. These objects were then expanded within the existing binary volume by iterative dilation with a sphere (diameter 3 voxels) until the relative increase in volume per step was smaller than 10%, to prevent growing of the cells into the canaliculi. The detected objects where masked, and the remaining volume was skeletonized using a custom vectorized implementation of parallel medial axis thinning in 3D (Lee, Kashyap et al. 1994), resulting in a one-voxel thick representation of the network of canaliculi connecting cell lacunae and blood vessels (Fig. 1a). Short branches resulting from image noise were pruned after skeletonization using a cutoff of 1 µm. The canalicular network was then converted into a weighted non-directed graph represented by nodes, edges and their adjacency matrix. Here, edge weight in the adjacency matrix corresponds to the length of the canaliculus between two intersections, or to the length of the shortest canaliculus if multiple connections exist. All image and network analysis was implemented in MATLAB (MATLAB 2013b, Mathworks) using the image processing toolbox.

## *2.4 Spatial network analysis*

The *total void fraction* is defined as the number of all foreground (lacunar or canalicular) voxels in the binary image divided by the total number of voxels. *Canalicular density (Ca.Dn)* is a measure for the density of cell processes within the matrix and is defined as the number of network voxels without lacunae divided by the total number of voxels without lacunae. The proximity of a matrix voxel to cell lacunae, $d_L$, and to the entire lacuno-canalicular network $d_{LC}$, was measured by calculating the 3D Euclidean distance transform of the background to the foreground voxels in the binarized images of lacunae only, or of the entire network including lacunae. The distance $d_{net}$ of any point within the network of canaliculi to reach the closest cell through the network was determined by iteratively tagging all canalicular voxels starting from cell lacunae. The (dimensionless) gain factor $G_{net}$ of transport efficiency between cells and matrix due to presence of the canalicular network was calculated as the *actual* distance to the next lacuna through the network divided by the *effective* distance if the transport through canaliculi is *k* times faster compared to passive diffusion through the bone matrix,

$$G_{net} = (d_{LC} + d_{net}) / (d_{LC} + 1/k * d_{net}),$$

averaged over all matrix voxels. Here, we define transport efficiency in the sense of how long signals and minerals travel between cells and matrix, assuming that travel time is proportional to distance. $G_{net}$ is a lower boundary for the actual time gain, since the scaling of transport time with distance is expected to be faster for directed transport through the network compared to passive diffusion through the matrix.

## *2.5 Topological network analysis*

The network topology is represented as a sparse connectivity matrix A, where $a_{ij}$ is the length of the shortest link between nodes i and j, or 0 if no such link exists. The size of the matrix corresponds to the number of nodes N and defines the graph size. The *edge density*, sometimes also called *connectivity*, is calculated as the number of existing edges E divided by the number of possible edges in an undirected graph, N(N-1)/2. The *degree*, or *weight*, of a node is the number of edges connected to it. Nodes with degree = 1 are end points of branches of the network, all other nodes have degree ≥ 3 by definition. Endpoints at the edge of the volume (< 3%) cannot be distinguished from cut-off canaliculi, but were kept to maintain consistency of the network. The *clustering coefficient (CC)* of a node, which is a measure for the local connectivity, is calculated by dividing the number of existing by the number of possible edges between all neighbors of a node. The *average shortest path (ASP)*, which measures the typical separation between two nodes in the network, is derived by taking the mean of all shortest paths between any two nodes. The *betweenness centrality* of a node is the fraction of all such shortest paths in the network running through that node. To compare the results for the OLCN with random networks, we simulated 50 equivalent Erdös-Renyi (ER) networks (identical size and density but randomly placed edges) for each sample, and estimated the values of the relevant parameters by averaging over them. Clustering coefficients, shortest paths, betweenness centrality and ER graphs were calculated using the Boost Graph Library for MATLAB by David F. Gleich (Gleich 2008).

## *2.6 Statistics*

All results are stated as mean ± 95% confidence interval unless stated otherwise. Differences between the two bone types and cutting planes were tested for by performing a two-way ANOVA (`anova2()` in MATLAB). In addition, individual differences between all four groups were assessed by multiple comparison testing (`multcompare()` in MATLAB) with Bonferroni correction. All means and p values are summarized in Tables S1-S2. Raw data and the scripts to perform the analysis and generate the figures are available for download[1].

---

[1] `https://github.com/phi-max/OCY_connectomics`

# 3 Results

## *3.1 Structural quantification*

### 3.1.1 Network density and void fraction

We compared the lacuno-canalicular network in fibrolamellar sheep bone and woven mouse bone from different locations and orientations. At a first glance, cell networks in fibrolamellar bone appear dense and well organized, with regularly spaced cell lacunae. In contrast, the lacunae in woven bone seem to be randomly arranged, and the canaliculi look more sparse and irregular (Fig. 1a). To quantify these observations, we assessed the density of the canalicular network expressed as length per volume, and related it to the total porosity of the tissue including both the canalicular network and the lacunae (Fig. 1c). The total void fraction is about three times as high in woven bone (0.16±0.05) compared to fibrolamellar bone (0.05±0.01) due to the presence of large voids that likely correspond to blood vessels. In contrast, the canalicular density (Ca.Dn) of the bone matrix in the fibrolamellar bone samples is 0.19±0.01 $\mu m/\mu m^3$ and therefore twice as high as compared to that in the woven bone samples (0.10±0.01 $\mu m/\mu m^3$). While the total void fraction depends on the sample region and exhibits high variability, the canalicular density is a local property and appears much more homogeneous across different samples from the same bone type. Overall, fibrolamellar bone has less total void volume but a denser canalicular network.

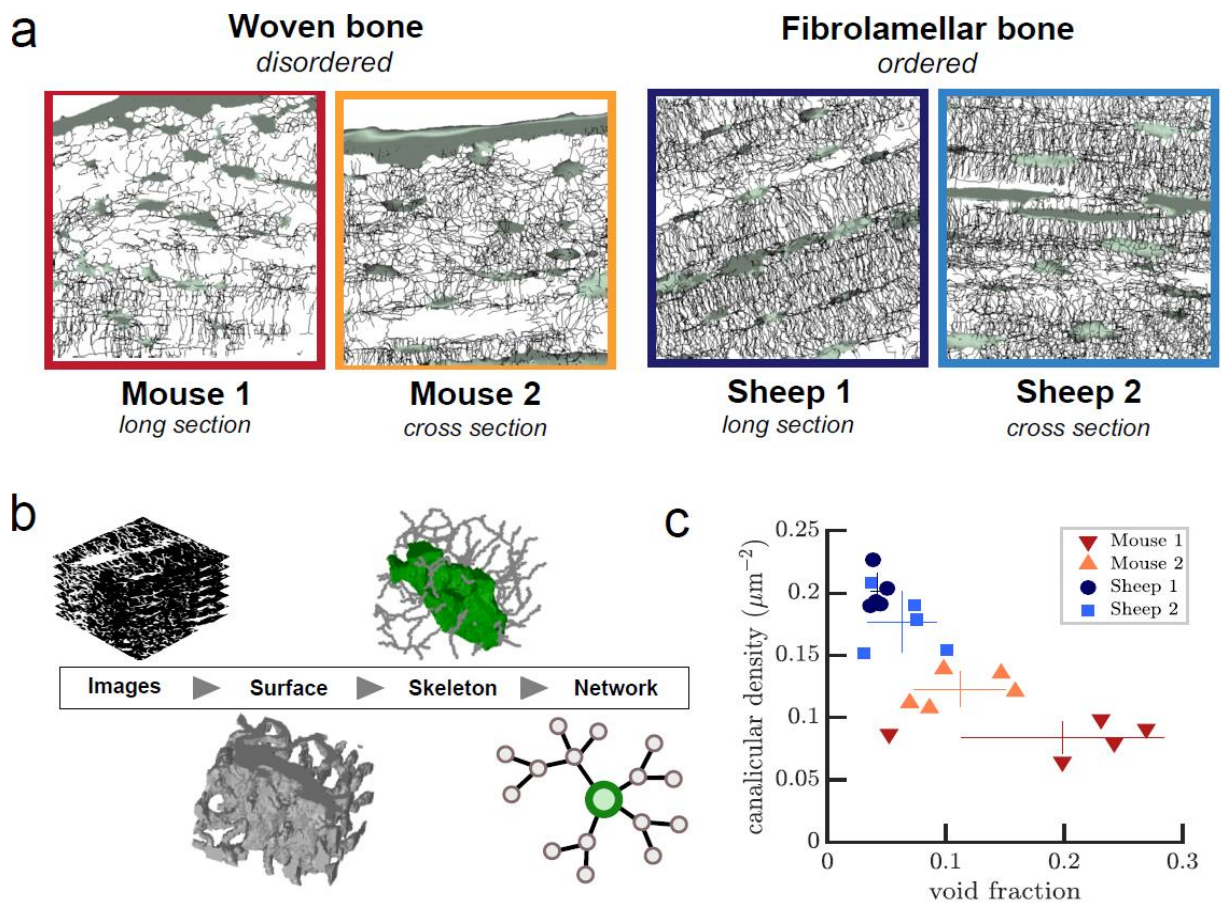

**Figure 1: Overview of the workflow and the resulting networks in different samples.** a) Appearance of the OLCN in the four different types of samples, from left to right: mouse woven bone cross section, mouse woven bone longitudinal section, ovine fibrolamellar bone cross section, and ovine fibrolamellar bone longitudinal section. Images are projections of the skeletonized data from one representative volume of each sample type. The total number of analyzed volumes was 20 (n = 5 per subgroup). Images of all samples are included as supplementary information. The field of view is always 100 µm x 100 µm. b) Image analysis workflow from left to right: raw confocal images, surface after thresholding, segmented skeleton, and network topology. c) Canalicular density over total void fraction for all individual samples (symbols), and means and standard error (lines) for each of the four sample types.

### 3.1.2 Proximity of matrix to bone surfaces

We next compared the accessibility of the bone matrix in the different samples by calculating the distance of the matrix from the closest lacunar or lacuno-canalicular surface. We found that the matrix is on average closer to lacuno-canalicular surfaces in fibrolamellar bone ($d_{LC}$ = 1.01±0.04 µm) compared to woven bone ($d_{LC}$ = 1.75±0.22 µm), but further from the next lacuna (Fig. 2a). Cumulative histograms in Fig. 2a show that in the two fibrolamellar bone groups, 93% and 96% of the matrix are within 2 µm from the network, whereas in the woven bone groups, only 64% and 77% are within 2 µm, respectively. Fig. 2b shows the 3D spatial distribution of the distance in representative volumes of each sample type. In woven bone of mice, the presence of large white and grey areas suggests that those regions of the tissue have only limited access to the lacuno-canalicular network.

### 3.1.3 Transport properties

We next quantified the distances within the canalicular network. Cell bodies and nuclei of osteocytes reside within the larger lacunae, whereas the canaliculi contain their thin cell processes. Since signals are processed in the cell bodies rather than in the protrusions, the distances that need to be overcome between cell bodies through the network are an important parameter for the intercellular communication between osteocytes. We found this distance to be on average smaller in woven bone ($d_{net}$ = 7.79±0.97 µm) compared to fibrolamellar bone ($d_{net}$ = 10.25±1.46 µm) (Fig. 2c). In other words, a single cell in fibrolamellar bone must span on average a larger matrix volume with its processes than a cell in woven bone.

In Fig. 2c, we compare the two previously quantified parameters, the distance $d_{LC}$ of the bone matrix from the closest lacuno-canalicular surface, and its effective distance $d_{net}$ from the closest cell lacuna through the network, since both distances are important for the accessibility of the matrix by cells. It is evident that, since fibrolamellar bone is more densely filled by a well-organized canalicular network, the distance to the network is smaller compared to that of woven bone. Interestingly, the variance of the two parameters is opposite in the two bone types: the variation of the distance to the next lacuno-canalicular surface is smaller in fibrolamellar bone, whereas the variance of the distance to the next cell lacuna is smaller in woven bone. The two sample groups *within* fibrolamellar and woven bone, respectively, were not different in either of the two parameters.

An important functional aspect of the canalicular network is that transport of molecules and signals is faster and therefore more efficient through the network than by diffusion through the bone matrix. We estimated the gain $G_{net}$ of efficiency due to the presence of the network as a function of the increase in transmission velocity through canaliculi ($v_{network}$) compared to through the matrix ($v_{matrix}$). An effective distance to the next cell was calculated as a function of the ratio $v_{network}/v_{matrix}$ and normalized against the actual physical distance (Fig. 2d, see methods). This corresponds to a decrease in the time necessary that a signaling molecule or mineral ion from the bone matrix reaches the next cell. For a velocity ratio of 1, $G_{net} = 1$ – in this case there would be no advantage of having a canalicular network for transport. In reality, the diffusion coefficient of small molecules through the collagen-apatite porosity of the bone matrix is estimated to be at least 100 times smaller than for free diffusion or load-enhanced fluid flow through the OLCN (Wang, Wang et al. 2005, Fritton and Weinbaum 2009, Marinozzi, Bini et al. 2014). For a ratio of k=100, the gain rises above 10 in fibrolamellar bone, and up to 6 in woven bone, and then saturates since it is now limited by diffusion to the closest canalicular surface. Therefore, at realistic values for the velocity ratio, the performance gain due to the presence of the canalicular network is more than twice as high in fibrolamellar compared to woven bone.

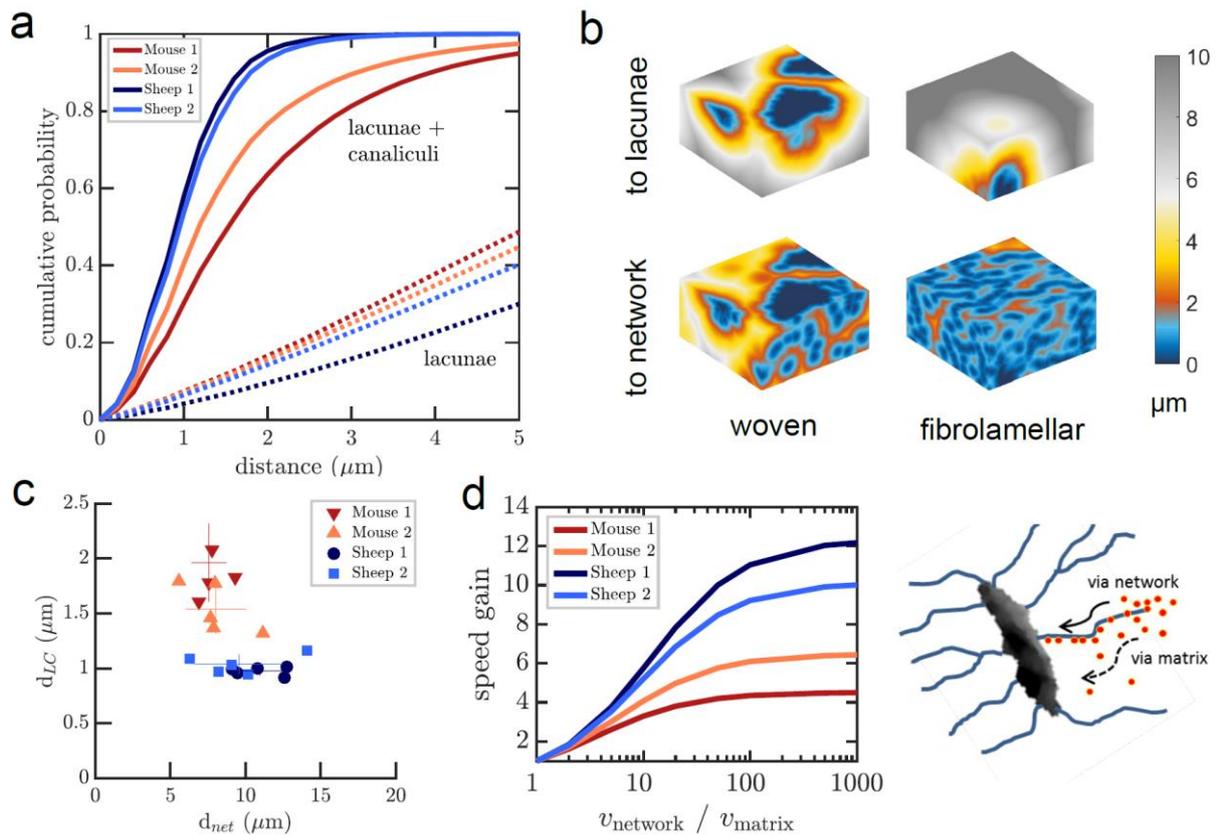

**Figure 2: Structural characterization reveals advantages of the more organized network in fibrolamellar bone compared to woven bone.** (a) Accessibility of the bone matrix from the lacuno-canalicular network (solid lines) and from lacunae (dashed lines). The fibrolamellar bone matrix is on average further away from the next lacuna, but closer to the lacuno-canalicular network and therefore to bone surfaces. (b) Examples of color-coded distance maps of the bone matrix from lacunae and the canalicular network for woven (left) and fibrolamellar bone (right). (c) Average distance of the network from the next lacuna ($d_{net}$), and of the matrix from the canalicular surface ($d_{LC}$). The network in fibrolamellar bone is further away from the next lacuna but on average closer to the matrix. (d) Effective gain of transport efficiency over the network compared to through the bone matrix. The higher the relative difference in velocity through the network, $v_{network}$, compared to diffusion through the matrix, $v_{matrix}$, the larger the benefit of having a denser lacuno-canalicular network.

### 3.2 Statistical properties

#### 3.2.1 Network size and edge density

The representation of the osteocyte lacuno-canalicular network as a graph enabled us to apply complex network analysis to determine statistical properties of the network, and to compare them between different bone types and with other biological networks. The most general properties of a graph are its size, defined as the number of nodes N, and its edge density, which is the fraction of existing edges over all possible edges. The size of the networks was larger in the more ordered fibrolamellar bone compared to woven bone (Fig. 3a). There was a clear difference between the two sample groups within both fibrolamellar and woven bone. This correlates with the apparent network density in the respective images (Fig. 1c). Since all samples had identical volume, differences in graph size N represent differences in the number of nodes per volume. The average node density was $4 \times 10^7$ per $mm^3$ in woven bone and $7 \times 10^7$ per $mm^3$ in fibrolamellar bone.

Remarkably, the network graphs for the different samples do not exhibit a size-independent edge density, as would be the case for a random graph (Fig. 3a). Rather, the universal invariant in the osteocyte network is the number of edges per node. When number of nodes (excluding cells and endpoints) and edges are plotted against each other, all examined samples regardless of species and bone type fall onto a single straight line, as shown in the inset to Fig. 3a. The number of edges per node averaged over all samples is $3.28 \pm 0.01$ if only nodes of degree $\geq 3$ are included, and there is no statistically significant difference between the four groups. This predicts an edge density that scales with the number of nodes as *$3.28 \times N^{-1}$*. This power law precisely fits the data in Fig. 3a. If also endpoints, i.e. nodes with degree 1, are included the average degree drops to 2.8 and becomes significantly different between sample groups due to the different percentage of endpoints (see below).

#### 3.2.2 Universal edge length and degree distributions

We next determined the cumulative statistical distributions of edge length and node degree, which are two common measures to classify and compare complex networks. Cumulative edge lengths were exponentially distributed in all examined networks (Fig. 3b) with a typical decay parameter of -2/3.

The decay parameter (or slope of the distribution in a semi-logarithmic survival plot, shown as a gray line in Fig. 3b) was not significantly different between any of the four sample groups.

Cumulative degree distributions for all four sample groups are shown in Fig. 3c. All four groups followed an exponential distribution with a decay (or slope in a semi-logarithmic plot, indicated by the gray line) around -4/3. Again, there was no significant difference in the decay parameter between the four groups. Taken together, all examined networks, despite their differences in appearance and density, universally follow the same exponential edge length and degree distributions.

In a spatial network, the "importance" of a node depends not only on the number of edges that connect to it, but also on the length, or "weight", of the edges. To quantify this, the "weighted degree" WD is calculated as the sum of the length of all edges connected to a node. The cumulative weighted degree distributions are shown in Fig. 3d. In this evaluation, cells were included and plotted together with all other nodes in one graph. The weighted degree distribution follows an exponential decay similar to the degree distribution, but with a long tail above about 200 µm due to the presence of cells. For comparison, the gray lines indicate an exponential decay of -1/2 and a power law decay with an exponent of 3/2, respectively.

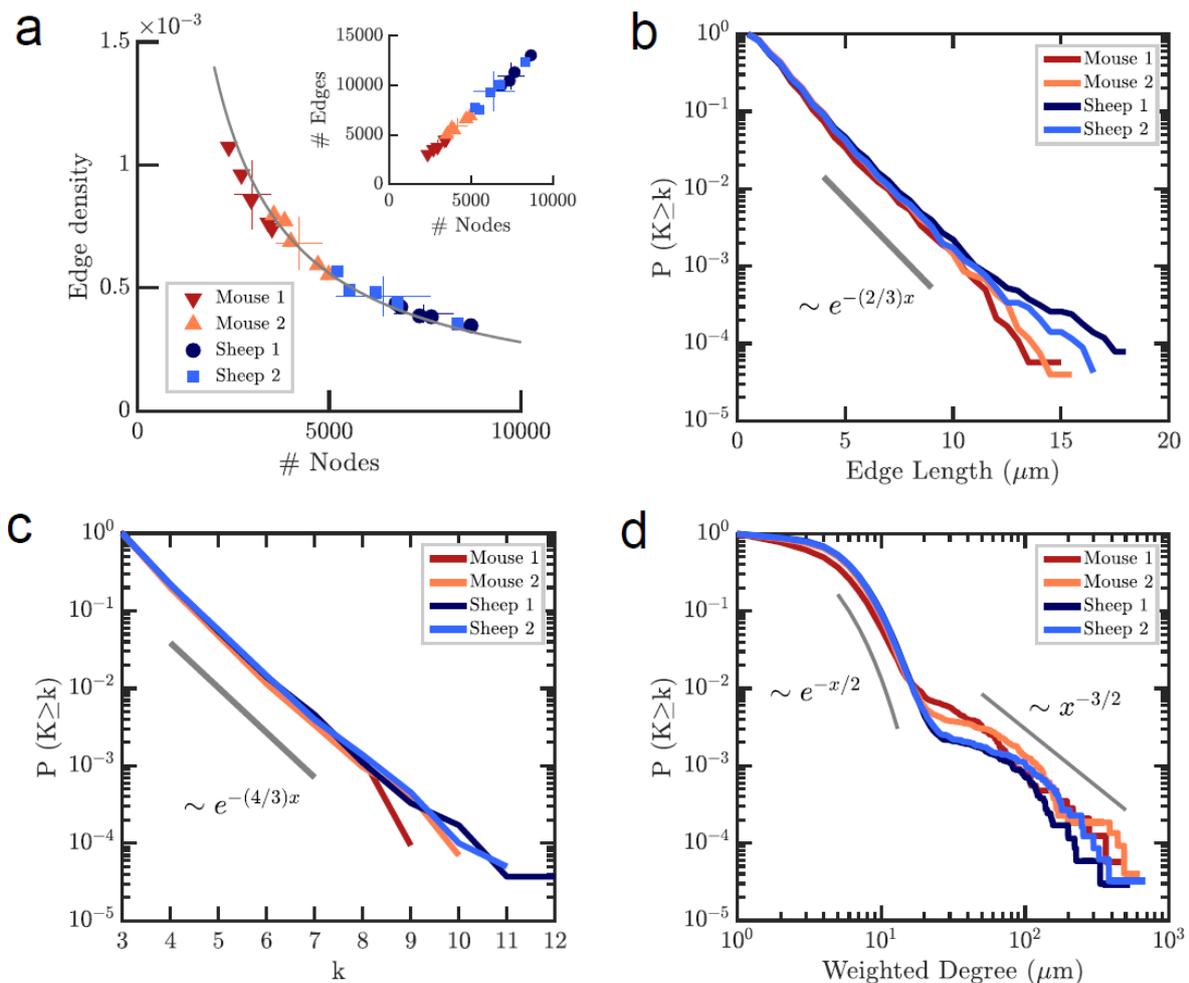

**Figure 3: Global statistical analysis and universal properties of the network topology.** (a) Edge density (number of existing over number of possible edges) versus number of nodes for individual samples. The edge density depends on network size, while the number of edges per node is identical in all samples (inset) and about 3.3. (b) Cumulative edge length distribution for all four sample types. The edge length distribution is in all cases exponential with a factor of -2/3 (gray line). (c) Cumulative degree distribution for all four sample types. Again, all four distributions are exponential with a decay of -4/3 (gray line), most nodes have degree 3. (d) Cumulative weighted degree distributions, taking into account the edge length and including cell lacunae as nodes. All four samples show an exponential distribution up to 200 μm with a decay of -1/2, and a power law tail at larger degrees with an exponent of -3/2.

## *3.3 Topology and connectivity*

### 3.3.1 Tree-like local topology

We next looked at the local topological organization of the network of interconnected canaliculi. Here, it is not only interesting to look at the number of neighboring nodes to which a node is connected (its degree), but also at the probability of connections *between* these neighboring nodes, expressed in the clustering coefficient (CC). A CC of one means that all neighbors of a node are connected to each other, while a clustering coefficient of zero means that there are no connections between neighbors of a node. The first case corresponds to a fully connected dense region of the network, whereas the latter case indicates a tree-like organization of this region of the network (Fig. 4a).

To characterize the local organization of the canalicular network, we classified nodes according to their clustering coefficient and their degree: nodes with a clustering coefficient larger than 0.5 (independent of their degree) were identified as „cluster nodes", whereas nodes with degree = 3 and clustering coefficient = 0 were defined as „tree nodes", indicating a dendritic-like organization of the network where a cell process branches into two sub-processes (Fig. 4 a,b). We found in all cases that the majority of nodes belonged to the category of tree nodes, suggesting that the network of cell processes is largely organized as a dendritic branching network. All investigated networks exhibited only a small number of cluster nodes. Woven bone samples had a higher percentage of cluster nodes (1.48±0.5 %) and a smaller percentage of tree nodes (43.0±3.9 %) compared to fibrolamellar bone (0.76±0.2 % and 53.5±1.7 %, respectively).

A third relevant class of vertices are the end points of branches, defined as nodes with degree = 1 that are not at the boundary of the volume (i.e. no cut-off canaliculi). The percentage of such nodes in the network is a measure for how often canaliculi end as "one-way roads" rather than being connected to other parts of the network. We found that on average, 29.8±4.2 % of all nodes in woven bone are such end points, whereas only 17.5±1.9 % of all nodes in fibrolamellar bone are end points (Fig. 4c).

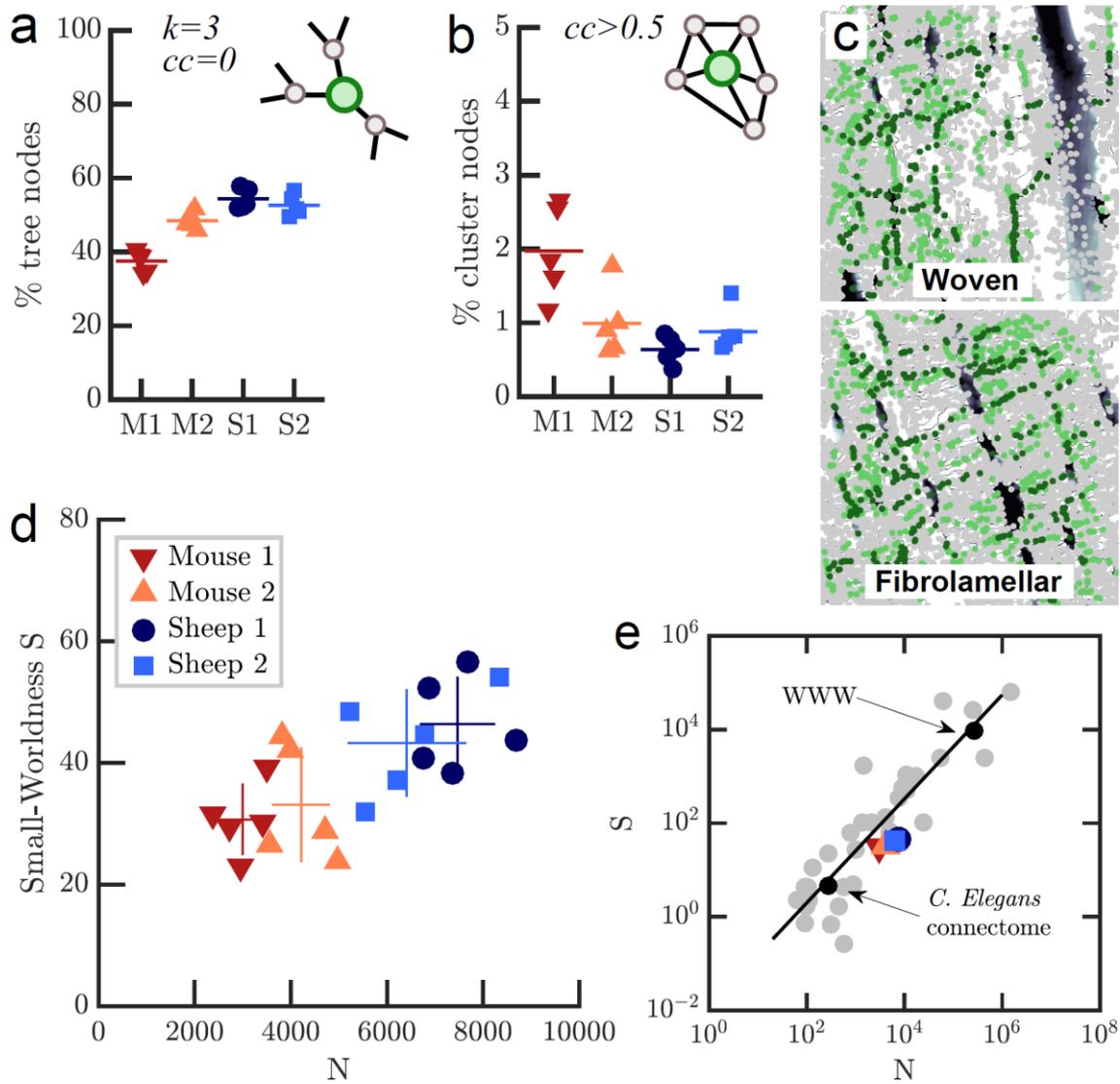

**Figure 4: Local topology and connectivity analysis reveals small-world properties.** (a) Percentage of tree-like nodes is higher in fibrolamellar than in woven bone. (b) Woven bone has a higher percentage of cluster nodes. (c) Spatial maps of betweenness centrality (number of shortest paths running through a node) reveals "highways" between cells and cell layers. Nodes with betweenness in the highest 15% are light green, while nodes within the highest 5% are dark green. (d) Small-Worldness S, defined as CC over average shortest path (ASP) relative to an equivalent Erdös-Renyi (ER) random network of same size and edge density, plotted over network size. Small-world networks have a similar ASP but larger CC compared to the equivalent ER network, resulting in S>1. (e) S for all four subgroups of osteocyte networks in comparison to other real world networks (data from Humphrey et al 2008). Black line is the fit of S over N for all data described in Humphrey et al 2008.

### 3.3.2 Average Shortest Path and Betweenness Centrality

Our investigation of the local topology showed that the network of osteocyte cell processes can be described as a dendritic branching network interspersed with a few high-CC nodes with many connections between neighbors. Woven bone had a higher percentage of highly clustered nodes compared to fibrolamellar bone. We next asked if these local differences in topology coincide with differences in global measures for topology and transport efficiency (e.g. for nutrients or chemical and mechanical signals). One possibility to express the efficiency of transport across a network is the average shortest path (ASP) from one node to any other in the network. The ASP averaged over all samples was 48.2 ± 3.7 µm, with no significant differences between sample groups despite the larger overall size of the network in fibrolamellar bone. For any given node, the *betweenness centrality*, defined as the fraction of shortest paths in the network running through that node, is a measure for the importance of that node. Fig. 4d shows two color-coded maps of the betweenness centrality of nodes for fibrolamellar and woven bone, respectively. It is immediately clear that in both bone types, high-centrality nodes align along preferential "highways" across the network through which many shortest paths are running.

### 3.3.3 Small-world properties

We next investigated if the osteocyte lacuna-canalicular system is a small world network. Small world networks are a class of networks where the average path length between any two nodes is much smaller than in a random network of same size and density, resulting in a more efficient communication (Amaral, Scala et al. 2000, Boccaletti, Latora et al. 2006). Many biological, technical and social networks have been found to belong to this class (Amaral, Scala et al. 2000, Eguiluz, Chialvo et al. 2005). Although there is no standard definition of a small world network, a number of practical measures for small-worldness have been proposed. The classical measure is that the average shortest path (ASP) is much smaller than in a random network of same size, but the clustering coefficient (CC) is similar (Amaral, Scala et al. 2000, Boccaletti, Latora et al. 2006). While the ASP was not significantly different between all subgroups, CC was slightly higher in woven bone than in fibrolamellar bone (p<0.05, Fig. 4g). A more rigid definition proposed by Humphrey et al. (Humphries and Gurney 2008) is small-worldness S, defined as the ratio of ASP and CC divided by the same ratio for a Erdös-Renyi random graph of the same size and edge density. We calculated this ratio for each of the 20 networks and found S to be 31.9±4.6 in woven bone and 44.8±5.0 for fibrolamellar bone (Fig. 4e). We observe a higher S for the larger fibrolamellar bone networks and a roughly linear relationship between N and S (Fig. 4e). We compared the network size and small-worldness S to those of >30 real small-world networks from (Humphries and Gurney 2008), including the C.Elegans neuronal connectome, electric circuit maps, and train networks (Fig. 4f). We find that the osteocyte network, with its intermediate size compared to the other networks, fits well into the linear relationship between S and N.

## 4  Discussion

In this study, we report an extensive quantification of the spatial and topological properties of the osteocyte lacuno-canalicular network in bones of different structural organization, i.e. woven bone from mouse and fibrolamellar bone from sheep. Woven bone has a higher overall void fraction but a smaller density of canaliculi compared to fibrolamellar bone (Fig. 1c). Concurrently, its matrix is on average further from the next canalicular surface (Fig. 2a), whereas the distance to the next lacuna both through the matrix as well as through the network is shorter (Fig. 2c). In woven bone, these mean distances exhibit more pronounced variation between samples compared to fibrolamellar bone (Fig. 2b). A smaller variation between samples suggests that this parameter is more controlled during development. Hence, the spacing and organization of the canaliculi appears to be more tightly regulated in fibrolamellar bone compared to woven bone. The higher degree of organization of the network in fibrolamellar bone is furthermore supported by the 2.5-fold higher gain in transport efficiency due to the presence of the network (Fig. 2d). Taken together, the spatial organization of the network in fibrolamellar bone is more efficient compared to woven bone in terms of controlling the bone matrix and accessing the stored mineral.

While the quantification of the spatial architecture of the OLCN revealed pronounced differences between all four sample groups, they exhibit surprisingly universal statistical properties on the level of network topology. The number of edges per node is 3.3 in all examined networks independent of size. Both edge length and node degree are exponentially distributed, with no statistically significant differences between the four sample groups. The presence of power-law tails of the weighted degree distribution including cell lacunae emphasizes the role of cells as "hubs" within the network.

Investigation of the local topology reveals that the network is mainly organized in a tree-like fashion, with only a small percentage of nodes with high clustering coefficient. This tree-like organization is the expected outcome of a dendritic outgrowth process with successively branching cell processes (Bonewald 2005, Buenzli and Sims 2015). The exponential degree distribution and presence of high-degree nodes in the network does not contradict such a branching process as the main organizing mechanism during growth, since high-degree nodes could also represent dense clusters of branching points (tree nodes) within a small volume below the optical imaging resolution (Wittig, Birkbak et al.

2016). All examined networks seem to exhibit similar local organization despite pronounced differences in the total network size.

The topological architecture of the network allows quantifying the efficiency of the network for information transfer. An important measure here is the average shortest path (ASP) between any two nodes in the network, and the number of such paths that run through a given node (betweenness centrality). We found that the shortest paths between nodes in the network run along a few "highways" on which most high-centrality nodes are clustered together. By comparing the ratio of ASP and CC to that of a random graph of the same size and density, we derive the small-worldness S and find it to be significantly higher in fibrolamellar compared to woven bone networks, and proportional to network size N. Proportionality between S and N has been reported for a large number of real-world networks as diverse as neuronal connectomes, transport and signaling networks with *$S \sim 0.023*N^{0.96}$* (Humphries and Gurney 2008). Despite a different proportionality observed for our data, this probably explains the observed difference in S for the different bone types. When we compare the average S and N for all samples to the data in Humphries and Gurney 2008, we find that the osteocyte network exhibits a similar S as other real-world complex networks of similar size (Fig. 4f). Since small-worldness is a measure for how efficiently the network is organized to transmit signals between distant points in the network, we conclude that the small-world like topology of the osteocyte network is a result of adaptation towards efficient organization.

How are these findings linked to the functions of the osteocyte network? As outlined in the introduction, the osteocyte lacuno-canalicular network is believed to be important for bone mechanosensing, signal transduction, and nutrient supply of osteocytes. In all three of these functional scenarios, the efficiency of the network in distributing molecules and signals between cells and throughout the matrix plays a central role. The optimal strategy would then be to access as much bone matrix volume as possible with as few cells as possible, while keeping the effective distance between cells through the network as short as possible. Our results show that fibrolamellar bone, despite containing fewer cells per volume, exhibits a more optimized organization of the canalicular network with respect to these functional criteria as compared to woven bone.

This advantage might be a consequence of the more controlled but slower growth of highly organized bone against a pre-existing endogenous scaffold, whereas woven bone is rapidly formed in a less controlled manner and in the absence of an organizing surface (Ferretti, Palumbo et al. 2002, Kerschnitzki, Wagermaier et al. 2011). The guidance by the scaffold surface results in long-ranged alignment of cells and extracellular matrix fibers during bone deposition. This oriented organization does not only improve the mechanical performance of bone, but serves as a blueprint for the osteocyte network. Cell processes align with the topographical cues of the surrounding matrix (Dunn and Heath 1976, Curtis and Wilkinson 1997) and direct the orientation of subsequently deposited extracellular matrix fibers (Wang, Jia et al. 2003, Lamers, Walboomers et al. 2010). This continuous feedback between cells and their environment gradually produces more coordinated and integrated higher-level structures. Together with the longer time available to the fibrolamellar network to grow and reorganize before the tissue structure is fixed due to mineralization, this allows for a more dense, well-organized and therefore efficient network. The higher degree of organization of the fibrolamellar OLCN together with the superior mechanical performance justifies why woven bone is gradually replaced by fibrolamellar bone, despite the high cost of resorbing and depositing bone tissue. Once the network is in place, the newly remodeled bone can more efficiently contribute to the existing mechano-sensory network and mineral depot of the skeleton.

Our results show that the organization of the osteocyte network can serve as a measure for bone quality (Seeman and Delmas 2006) due to its direct functional relevance. The quantification developed here may be useful in assessing bone quality during physiological development or pathological conditions of age, disease and pharmaceutical intervention, complementary to existing parameters such as bone mineral density. Although we did not apply our analysis to compare healthy to diseased bone, our choice of different bone types reflecting different degrees of organization demonstrates the potential of our method to quantify differences in efficiency. Pathological deviations in matrix protein structure, matrix deposition, or bone homeostasis likely result in a less efficiently organized network. Conversely, deficits in network organization lead to less efficient mineral exchange, impaired mechanotransduction and overall reduced osteocyte viability due to limited nutrient supply. By quantifying the parameters of the OLCN as introduced here, such two-way interactions between network organization and bone quality can be easily assessed by comparing healthy and diseased bone samples.

While the density and spatial organization of the network differ between bone types, other properties of the network, such as edge length distribution, degree distribution, and the tree-like topology are surprisingly universal despite the different growth modalities. This shows that the same cellular mechanisms is at work during the outgrowth of cell processes and the formation and pruning of cell-cell contacts, independent of bone type and species. It also suggests that these mechanisms operate in a narrow parameter range that results in an optimized topology with regards to node density and edge length. For example, a certain fraction of high-degree nodes, or closely spaced 3-nodes, might be ideal for sensing damage or integrating signals from remote branches of the network.

One fundamental challenge in biology is to understand how individual cells come together and form higher-order structures that perform complex functions. Multicellular networks such as the OLCN are a common example of such emergent self-organization. Complex network theory offers a well-established framework to robustly quantify and compare such structures using measures from statistical physics and graph theory. Our results provide insights into which principles might be at work when individual bone cells self-organize into a large, interconnected network during bone formation and mineralization. The workflow that was used here (volume imaging followed by segmentation and conversion into a graph) is broadly applicable to many biological network-like structures at different length scales. One remarkable outcome of our analysis is that the osteocyte network shares many characteristics with other multicellular networks, such as neuronal networks, and with higher-order network structures such as the vascular system or the airways (Eguiluz, Chialvo et al. 2005, Bullmore and Sporns 2009, Blinder, Tsai et al. 2013, Herriges and Morrisey 2014). The analogy between the OLCN and neuronal networks has previously inspired speculations about the potential signal processing power of the whole skeleton (Turner, Robling et al. 2002). The approach presented here would allow for a more in-depth quantitative comparative analysis between different networks, potentially revealing unexpected similarities and universal ordering principles.

## 5 Conclusion

We developed a strategy for the quantification of the architecture of the osteocyte network using different bone types from mouse and sheep for the proof of principle. Based on these data, we reported the first detailed characterization of its topological and statistical properties on the cellular level. We defined a number of robust, quantitative measures that are derived from the theory of complex networks, and used these measures to gain insights into how efficient the network is organized with regard to intercellular transport and communication. Our analysis shows that, on the canalicular level, highly organized fibrolamellar bone from sheep, which grows slower but in a more controlled way on top of a template surface, is less interconnected, but more efficiently organized than woven mouse bone, which grows faster but in the absence of an organizing surface. Despite pronounced differences at the tissue level, the topological architecture of the osteocyte canalicular network at the subcellular level is identical across sample types, suggesting a universal growth mechanism during network formation. Our method could be useful for comparing quantitatively the quality of the network of canaliculi in bone of different tissue and individuum age, loading condition and disease state. The results presented here provide insight in which principles might be at work when individual bone cells self-organize into a large, interconnected network during bone formation and mineralization.

## 6 Acknowledgement

We thank H. Schell from the Julius Wolff Institute of Charité, Berlin for providing ovine bone samples collected in the SFB 760, and the Institute of Biology of the University of Potsdam for providing the murine bone samples.

**Table S1**

Results for the four individual sample groups, and average ± CI over bone type and over all samples.

| | void fraction | Ca.Dn | dist$_{LC}$ | dist$_{net}$ | G | N/V | <k> |
|---|---|---|---|---|---|---|---|
| unit | - | $\mu m^{-2}$ | $\mu m$ | $\mu m$ | - | $\mu m^{-3}$ | - |
| M1 | 0.20 ± 0.08 | 0.08 ± 0.01 | 1.96 ± 0.31 | 7.55 ± 1.01 | 4.35 ± 0.96 | 0.03 ± 0.00 | 3.28 ± 0.02 |
| M2 | 0.11 ± 0.03 | 0.12 ± 0.01 | 1.54 ± 0.20 | 8.03 ± 1.75 | 6.09 ± 1.89 | 0.04 ± 0.01 | 3.26 ± 0.01 |
| S1 | 0.04 ± 0.00 | 0.20 ± 0.01 | 0.97 ± 0.03 | 10.94 ± 1.50 | 11.04 ± 1.21 | 0.07 ± 0.01 | 3.29 ± 0.01 |
| S2 | 0.06 ± 0.03 | 0.18 ± 0.02 | 1.04 ± 0.08 | 9.55 ± 2.54 | 9.21 ± 2.04 | 0.06 ± 0.01 | 3.29 ± 0.03 |
| M | 0.16 ± 0.05 | 0.10 ± 0.01 | 1.75 ± 0.22 | 7.79 ± 0.97 | 5.22 ± 1.15 | 0.04 ± 0.01 | 3.27 ± 0.01 |
| S | 0.05 ± 0.01 | 0.19 ± 0.01 | 1.01 ± 0.04 | 10.25 ± 1.46 | 10.13 ± 1.27 | 0.07 ± 0.01 | 3.29 ± 0.01 |
| all | 0.10 ± 0.03 | 0.15 ± 0.02 | 1.38 ± 0.20 | 9.02 ± 1.02 | 7.68 ± 1.38 | 0.05 ± 0.01 | 3.28 ± 0.01 |

| | len. dist. | deg. dist. | % tree nd | % clust nd | % end pt | <ASP> | <CC> | S |
|---|---|---|---|---|---|---|---|---|
| unit | - | - | - | - | - | $\mu m$ | - | - |
| M1 | -0.68 ± 0.05 | -1.35 ± 0.09 | 37.51 ± 2.49 | 1.97 ± 0.55 | 36.06 ± 1.88 | 47.88 ± 8.12 | 0.05 ± 0.01 | 30.70 ± 5.12 |
| M2 | -0.70 ± 0.05 | -1.41 ± 0.07 | 48.51 ± 1.97 | 0.99 ± 0.41 | 23.59 ± 1.46 | 50.43 ± 8.78 | 0.04 ± 0.01 | 33.11 ± 8.25 |
| S1 | -0.65 ± 0.03 | -1.36 ± 0.08 | 54.35 ± 2.38 | 0.64 ± 0.16 | 16.52 ± 2.01 | 49.60 ± 4.37 | 0.04 ± 0.00 | 46.38 ± 6.82 |
| S2 | -0.70 ± 0.06 | -1.34 ± 0.08 | 52.68 ± 2.47 | 0.88 ± 0.26 | 18.50 ± 3.17 | 44.93 ± 8.95 | 0.04 ± 0.00 | 43.29 ± 7.73 |
| M | -0.69 ± 0.03 | -1.38 ± 0.06 | 43.01 ± 3.89 | 1.48 ± 0.45 | 29.83 ± 4.22 | 49.15 ± 5.70 | 0.05 ± 0.00 | 31.91 ± 4.64 |
| S | -0.67 ± 0.04 | -1.35 ± 0.05 | 53.51 ± 1.71 | 0.76 ± 0.16 | 17.51 ± 1.88 | 47.27 ± 4.94 | 0.04 ± 0.00 | 44.84 ± 4.97 |
| all | -0.68 ± 0.02 | -1.36 ± 0.04 | 48.26 ± 3.14 | 1.12 ± 0.29 | 23.67 ± 3.57 | 48.21 ± 3.69 | 0.04 ± 0.00 | 38.37 ± 4.40 |

**Table S2**

Results of statistical testing for differences between sample groups using multiple testing with Bonferroni correction (first six lines), and for testing for differences between cutting planes p(1-2) and between bone types p(M-S) using two-way ANOVA. The p-values suggesting a difference between cutting planes and bone type for the respective parameter at 95% confidence level (p<0.05) are highlighted in red.

| | void fraction | Ca.Dn | $dist_{LC}$ | $dist_{net}$ | G | N/V | <k> |
|---|---|---|---|---|---|---|---|
| M1-M2 | 0.079 | 0.016 | 0.043 | 1 | 0.89 | 0.19 | 1 |
| M1-S1 | 0.00075 | 6.3e-08 | 1.1e-05 | 0.11 | 0.00015 | 1.3e-06 | 1 |
| M1-S2 | 0.003 | 1.5e-06 | 2.6e-05 | 0.85 | 0.0038 | 3.8e-05 | 1 |
| M2-S1 | 0.24 | 1.4e-05 | 0.0042 | 0.23 | 0.0032 | 7,00E-05 | 0.25 |
| M2-S2 | 0.84 | 0.00087 | 0.012 | 1 | 0.091 | 0.0038 | 0.27 |
| S1-S2 | 1 | 0.25 | 1 | 1 | 0.79 | 0.37 | 1 |
| p(1-2) | 0.025 | 0.18 | 0.013 | 0.28 | 0.08 | 0.093 | 0.38 |
| p(M-S) | 2.7e-05 | 1.6e-07 | 3.5e-07 | 0.023 | 3.6e-05 | 1.8e-07 | 0.02 |

| | len. dist. | deg. dist. | % tree nd | % clust nd | % end pt | <ASP> | <CC> | S |
|---|---|---|---|---|---|---|---|---|
| M1-M2 | 1 | 1 | 4.2e-05 | 0.014 | 4.8e-06 | 1 | 1 | 1 |
| M1-S1 | 1 | 1 | 1.7e-07 | 0.00092 | 9.9e-09 | 1 | 0.075 | 0.044 |
| M1-S2 | 1 | 1 | 7,00E-07 | 0.0058 | 4.6e-08 | 1 | 0.24 | 0.15 |
| M2-S1 | 0.75 | 1 | 0.019 | 1 | 0.0026 | 1 | 0.54 | 0.12 |
| M2-S2 | 1 | 1 | 0.15 | 1 | 0.035 | 1 | 1 | 0.38 |
| S1-S2 | 1 | 1 | 1 | 1 | 1 | 1 | 1 | 1 |
| p(1-2) | 0.47 | 0.9 | 0.026 | 0.031 | 0.041 | 0.14 | 0.63 | 0.26 |
| p(M-S) | 0.36 | 0.42 | 5.9e-06 | 0.00034 | 0,000005 | 0.61 | 0.013 | 0.004 |